\documentclass[prb,twocolumn,showpacs]{revtex4}
\usepackage{graphicx,epsfig,color}

\newcommand{\is}{\it {s}}
\newcommand{\ih}{\it {h}}
\newcommand{\ip}{\it {p}}
\newcommand{\ic}{\it {c}}
\newcommand{\ib}{\it {b}}

\newcommand{\hp}{\it {h-p}}

\begin{document}

\title{Cs adsorption on Si(001) surface: ab initio study}

\author{R.~Shaltaf$\,^{a,}$\footnote{Corresponding
author: \indent e-mail: shaltaf@pcpm.ucl.ac.be}$\footnote{Present
address: Unit{\'e} Physico-Chimie et de Physique des Mat{\`e}riaux,
Universit{\'e} catholique de Louvain, Place Croix du Sud, 1 B-1348
Louvain-la-Neuve, Belgique}$, E.~Mete$\,^b$, and
\c{S}.~Ellialt{\i}o\u{g}lu$\,^a$}

\affiliation{$^a$Department of Physics, Middle East Technical University,
Ankara 06531, Turkey \\
$^b$Physics Department, University of Bal{\i}kesir, 10100
Bal{\i}kesir, Turkey}

\date{\today}

\begin{abstract}

First-principles calculations using density functional theory based
on norm-conserving pseudopotentials have been performed to
investigate the Cs adsorption on the Si(001) surface for 0.5 and 1
ML coverages. We found that the saturation coverage corresponds to 1
ML adsorption with two Cs atoms occupying the double layer model
sites. While the energy bands spectra of the 0.5 ML covered surface
is of metallic nature, we found that 1 ML Cs adlayer leads to a
semiconducting surface. The results for the electronic behavior and
surface work function suggest that adsorption of Cs takes place via
polarized covalent bonding. The nature of Cs-Si bond shows a hybrid
character of $s-p$ type implying a limited charge transfer from the
adsorbate to the substrate.

\end{abstract}


\maketitle

\section{Introduction}

The adsorption of alkali metals on semiconductor surfaces has been
the topic of interest for several decades. The experimental and
theoretical studies were motivated by various technological
applications as in the low-temperature and low-field electron
emission sources and as for the catalyzers of silicon dioxide growth
on silicon surfaces. In the latter case Cs might be a better choice
because the adsorption of Cs atoms on Si surface alters the Si
surface structure only a little, which enables them to be desorbed
from the Si surface very easily.

First experiments on Cs/Si(001) system were performed by
Goldstein\cite{Goldstein} and Levine\cite{Levine} in the early
seventies. Using low-energy-electron-diffraction (LEED) and
Auger-electron-spectroscopy (AES) they proposed a model in which Cs
atoms are adsorbed on pedestal sites forming single Cs chains
running along the dimer rows, hence the saturation coverage was
believed to be 0.5 ML. Based on this model the electronic properties
of the covered surface will have a metallic nature as can be
predicted from simple electron counting. This model was generally
accepted in the proceeding years until the x-ray photoelectron
diffraction (XPD) results of Abukawa and Kono\cite{Abukawa} revealed
that at saturation coverage, alkali adsorbent atoms, like K and Cs,
will form two sets of arrays (a double layer), with a vertical
separation of $\sim$~1 \AA. Abukawa and Kono proposed a model in
which the atoms get adsorbed simultaneously on two different sites,
namely hollow and pedestal sites. The electronic properties of
Cs/Si(001) as investigated by Enta et al.\cite{Enta1} using
angle-resolved ultraviolet photoemission spectroscopy (ARUPS) were
consistent with the double layer model. Their results for surface
state dispersion showed a semiconducting surface which agrees with
the double layer model. Smith et al.\cite{smith} estimated using
medium energy ion scattering in conjunction with AES and LEED a
saturation coverage of $0.97 \pm 0.05$, a result which also supports
the double layer model.

Even though Abukawa model\cite{Abukawa} was adopted and supported in
many other works
\cite{Kim91,Lin,Mangat,Chao1,Chao2,Chao3,Benemanskaya,Hamamatsu},
the saturation coverage and adsorption sites of Cs/Si(001) still
stay as the matter of debate. In a recent experimental work using He
Rutherford backscattering, Sherman et al.\cite{Sherman1} found that
saturation can not exceed 0.5 ML supporting the Levine model, they
rejected the existence of DML (double monolayer) model. Kim et
al.\cite{Kim} also arrived at the same conclusion based on their
coaxial impact collision ion scattering spectroscopy. Meyerheim et
al.,\cite{Meyerheim} however, accepted the saturation coverage being
1 ML, but they suggested that adsorption on bridge (mid-point above
the Si dimer) and hollow sites might be more stable than the Abukawa
model.

The atomic reconstruction of covered surface, the bonding between
the Cs atom and the Si surface, as well as the electronic properties
of the covered surface still present a matter of dispute. For
instance, it was suggested by various
experiments\cite{Abukawa98,Meyerheim} that adsorption of Cs
symmetrizes the dimers. However, Chao et
al.,\cite{Chao1,Chao2,Chao3}  argued that the dimer structure is
still asymmetric. On the other hand, using ARUPS\cite{Enta1},
extended x-ray absorption fine structure
spectroscopy\cite{Kim91,Mangat} (EXAFS), and threshold photoemission
spectroscopy\cite{Benemanskaya} results it was concluded that the
adsorption of Cs atoms on Si(001) surface takes place via covalent
bonding, moreover, partially ionic bonding was suggested by
synchrotron radiation photoemission by Lin et al.,\cite{Lin} and
even further, a pure ionic bonding was proposed by the low energy
D-scattering results of Souda et al.\cite{Souda}

In this work, we employed ab initio total energy-pseudopotential
method to study the adsorption of Cs on Si(001) surface at the
previously proposed saturation coverages. We investigated the
structural and electronic properties as well as the work function of
the covered surface to get a better understanding of this system.

\section{Method}

We used pseudopotential method based on density functional theory in
the local density approximation. The self consistent norm conserving
pseudopotentials of silicon and hydrogen were generated by using the
Hammann scheme\cite{Ham-89} that is included in the fhi98PP
package.\cite{FS-99} Nonlinear core corrections were included in the
norm-conserving pseudopotential of Cs generated by using the
Troullier-Martins scheme.\cite{TM} In order to solve the Kohn-Sham
equations, conjugate gradients minimization method\cite{Pay} was
employed as implemented by the ABINIT code.\cite{abinit,Gonze} The
exchange-correlation effects were taken into account within the
Perdew-Wang scheme\cite{PW-92} as parameterized by Ceperley and
Alder.\cite{CeperleyAlder}

The surface unit cell (SUC) includes a slab with 8 layers of Si
atoms and a vacuum region equal to about 9 {\AA} in thickness.
Single-particle wave functions were expanded using a plane wave
basis up to a kinetic energy cut-off equal to 16 Ry. The Brillouin
zone integration was performed using 16 special $\vec k$-points that
are sampled with the Monkhorst-Pack scheme.\cite{MP-76}

We used the 2$\times$1 unit cell for the clean Si(001) surface in
our calculations, which is obtained by using the above parameters,
and having the lowest two layers (out of 8) kept fixed in their
ideal bulk positions while all the remaining substrate atoms were
allowed to relax into their minimum energy positions.

\begin{figure}[htb]
\epsfig{file=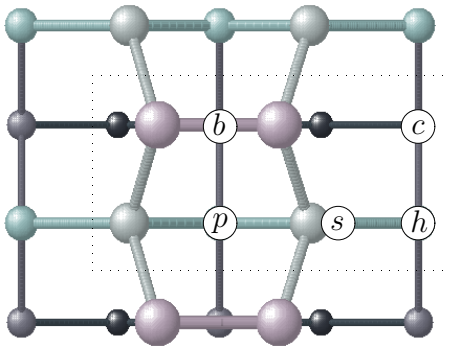,width=6cm,clip=true} \caption{Schematics of the
adsorption sites for Cs atom on Si(001) surface. Surface unit cell
is indicated by dashed box. The symbols stand for: $b$=bridge,
$p$=pedestal, $h$=hollow, $s$=shallow and $c$=cave. (The dimers are
shown symmetric here for visual convenience)\label{figure1}}
\end{figure}

The resultant relaxed structure has asymmetric dimers with a dimer
length of 2.27 {\AA} and a tilt angle of $18.1^\circ$. Our result for
clean Si(001) is in agreement with the theoretical results of
Ramstad et al.\cite{Ram} who found a dimer length of 2.26 {\AA} with a
tilting angle being 18.3$^\circ$.

We performed separate self-consistent calculations to determine
surface work function and formation energy using a symmetrical slab
having a thickness equal to 10 Si layers and being separated from
each other with a vacuum region of $\sim$15 \AA. After relaxing the
clean surface into its minimum energy geometry we froze the 4 Si
atoms which are located in the middle of this slab. We also repeated
geometry optimizations using symmetrical slab for some of the
coverage models as was done in the case of hydrogenated surface.
These calculations produced equivalent results.

\section{Results and Discussion}

We have studied the adsorption of Cs atoms on the Si(001) surface
for 0.5 ML and 1 ML coverages, starting with the reconstructed
$(2\times$1) surface unit cell. We have chosen five different sites
for the adsorption, namely, the cave, hollow, pedestal, bridge and
the shallow sites. The cave site ({\ic}) is located above the fourth
layer Si, hollow ({\ih}) and pedestal ({\ip}) sites are above the
third layer Si, shallow site ({\is}) is above the second layer Si
between the two Si dimers, and the bridge site ({\ib}) is located
above the dimer as indicated in Figure~\ref{figure1}. The adatom and
the substrate were then taken to their minimum energy configurations
by performing structural optimization using the
Broyden-Fletcher-Goldfarb-Shanno method\cite{BFGS} until the force
on each atom reduces to a value less than 25 meV/\AA.

\subsection{Structural Properties}

We have found that adsorption of Cs on {\ih} site is the most stable
one with an adsorption energy per adatom, $E_{\rm ad}$, equal to
2.46 eV, which is the negative of the binding energy of the adatom,
and defined by the expression
\begin{equation} n \, E_{ad}=(E_{\rm Si(001)}+n \, E_{\rm Cs})-E_{\rm
Cs/Si(001)} .
\end{equation} Here, $E_{\rm Cs/Si(001)}$ is the total energy of
the covered surface, $E_{\rm Si(001)}$ is the total energy of the
clean surface, $n$ is the number of Cs adatoms in a surface unit
cell and $E_{\rm Cs}$ is the total energy of a single Cs atom with
the spin polarization obtained in a separate ab initio calculation
using the same pseudopotential and the same kinetic energy cut off,
but in a larger unit cell of size $\sim$26 \AA.
\begin{table}[htbp]
\caption{The structural parameters for most of the stable adsorption
sites for two different coverages $\Theta$. The dimer lengths,
heights of the adatoms with respect to the second silicon monolayer
(all in \AA) and the adsorption energies ($E_{\rm ad}$ in eV) for
these cases are presented. The quantities in parenthesis are tilt
angles (in degrees) of the corresponding dimers.\label{table1}}
\vspace{1.5mm}
\begin{ruledtabular}
\begin{tabular}{l|c|c|c|r}
$\Theta$ \hspace{15pt} & \hspace{5pt} Model \hspace{5pt} &
\hspace{15pt} dimer \hspace{15pt} & \hspace{15pt} d$_{\perp}$
\hspace{15pt} & \hspace{15pt} E$_{\rm ad}$ \\[1mm]
\hline\hline &&&& \\[-2mm] 0.5 %1/2 $\frac{1}{2}$
& {\ih} & 2.34 ~~(0.8) & 2.75 & 2.46 \\[2mm]
& {\ip} & 2.39 ~~(8.7) & 3.75 & 2.02 \\[2mm]
& {\ic} & 2.31 ~~(7.5) & 3.15 & 1.89 \\[2mm]
& {\ib} & 2.37 ~(11.1) & 4.30 & 1.46 \\[1mm]
\hline &&&& \\[-2mm] 1
& {\ih}-{\ip} & 2.46 & 2.67 \hspace{2pt} 3.86 & 2.39 \\[2mm]
& {\ih}-{\ib} & 2.46 & 2.66 \hspace{2pt} 4.50 & 2.18 \\
\end{tabular}
\end{ruledtabular}
\end{table}

The adsorption of Cs atom on this site symmetrizes the dimers with
resulting dimer lengths equal to 2.34 {\AA}. Other configurations of
adsorption on {\ip}, {\ic} and {\ib} sites were found less stable
with adsorption energies of 2.02, 1.89 and 1.46 eV, respectively,
and with dimers still tilted in various angles as listed in Table
\ref{table1}. Finally, the adsorption on {\is} site was not found as
a well defined local minimum with the adatom migrating into a
neighboring {\ih} site. Our results suggest that if the saturation
coverage would exist then it must contain {\ih} site as being the
most probable adsorption site. Therefore, Levine's model appears to
be obsolete which assumes the adsorption of Cs atom on a {\ip} site.
On the other hand, our agreement is merely for the adsorption site
with Kim et al.\cite{Kim} who found that the saturation coverage
corresponds to 0.5 ML with a single Cs atom being adsorbed on {\ih}
site only. The height of Cs atom from the second silicon layer is
found to be 2.75 {\AA}, which is smaller than their experimental
value\cite{Kim} of 3.18 $\pm$ 0.05 {\AA}. The distance between Cs atom
and the nearest Si atom was found to be 3.58 {\AA} which is close to
their experimental value \cite{Kim} of 3.71 $\pm$ 0.05 {\AA}. Hashizume
et al.,\cite{Hashizume} using field ion scanning tunneling
microscopy, studied the adsorption at very low coverage, and
suggested that adsorption takes place at the off-center hollow site
(a non-symmetric site between {\ih} and {\is} in our notation). This
site has been also accepted by Gorelik et al.\cite{Gorelik} who made
use of local surface photo voltage (SPV), current imaging, and also
scanning tunneling microscopy, however, it has been rejected by a
recent experimental work by Park et al.,\cite{Park} who found rather
the {\ih} site as the most stable one. While the off-center site
might be stable for low coverages as reported by the
references\cite{Hashizume,Gorelik} we did not find any theoretical
evidence for this at the coverages studied in this work. Our aim
being to study the cases close to the previously reported saturation
coverages, this off-center site was also checked as the starting
point for the 0.5 ML adsorption case and was verified that the Cs
adatom migrates to the closest {\ih} site.

\begin{figure}[htb]
\epsfig{file=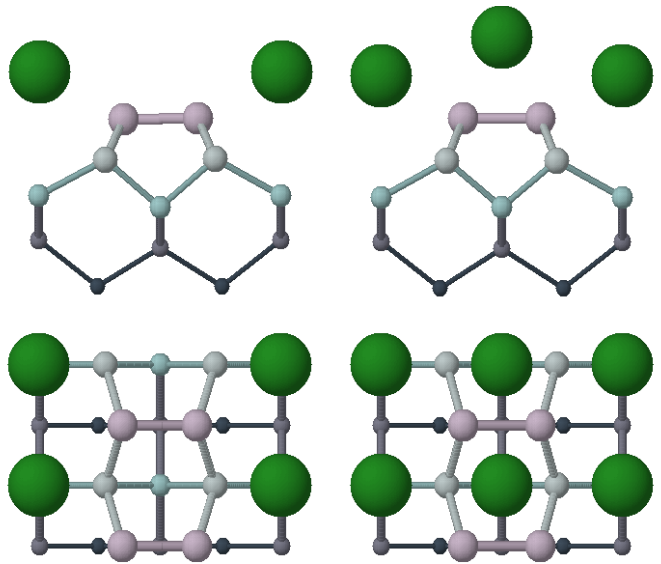,width=7.4cm,clip=true,angle=90} \caption{Cs
overlayer adsorbed on Si(001) surface (a) for 0.5 ML coverage on
{\ih} site and (b) for 1 ML coverage on {\hp} site. Only the upper
part of the slab is shown here. \label{figure2}}\end{figure}

For the full coverage we have considered only the combinations that
included {\ih} site as the starting configurations. We have four
different combinations from which we found that the Abukawa model,
with Cs atoms adsorbed at the {\hp} sites, was found as the most
stable one with an adsorption energy of 2.39 eV. The heights of the
Cs atoms as measured from the second Si ML, were 2.67 {\AA} and 3.86 {\AA}
for {\ih} and {\ip} sites, respectively. The double layer height
difference corresponds to 1.19 {\AA}  which is in excellent agreement
with the experimental value\cite{Abukawa} of 1.2 {\AA}. Meyerheim et
al.\cite{Meyerheim} using surface x-ray diffraction investigated the
atomic structure of Cs/Si(001), and found that the saturation
coverage corresponds to the combination of two adsorption sites
{\ib} and {\ih}. According to our results, even though this model
was found stable, it is less stable than the Abukawa model by 0.2 eV
per Cs adatom. Figure 2 shows the most stable configurations for
each coverage.
\begin{table}[htb]
\caption{The structural parameters (in {\AA}) for {\hp} model versus the
previously reported experimental work. The double layer height
differences, $\Delta_\bot$, are also presented.\label{table2}}
\vspace{1.5mm}
\begin{ruledtabular}
\begin{tabular}{c|c|c|c|c}
& \hspace{5pt} dimer \hspace{5pt}& \hspace{5pt} Cs-Si$_1$
\hspace{5pt}&
\hspace{5pt} Cs-Si$_2$ \hspace{5pt}& \hspace{5pt} $\Delta_\bot$ \\[1mm]
\hline\hline &&&&\\[-3mm]
This work & 2.46 & 3.41 & 3.49 & 1.19 \\[2mm]
EXAFS~\cite{Mangat}&$2.48\pm0.06$&$3.57\pm0.06$&&\\[2mm]
Tensor LEED~\cite{Hamamatsu} & 2.50 & 3.72 & 3.94 & 0.66 \\[2mm]
XPD~\cite{Abukawa}& -- & -- & -- & $1.2\pm0.1$\\
\end{tabular}
\end{ruledtabular}
\end{table}

In Table~\ref{table2} we tabulate our calculated structural
parameters for the {\hp} case along with the experimental values.
Our calculated bond lengths for Cs-Si bonds, being 3.41 {\AA} and 3.49
{\AA}, were in good agreement with 3.57 {\AA} obtained from
EXAFS\cite{Mangat} and somewhat smaller than the tensor LEED
values\cite{Hamamatsu} of 3.72 {\AA} and 3.94 {\AA}. We did not find that
the adsorption of Cs atoms on the surface leads to significant
alteration of the substrate. For the case of double layer adsorption
we found that the major changes in the substrate take place in the
dimer layer (Si first layer), where the upper atom in dimer gets
pushed down by a distance of 0.09 {\AA} while the lower atom is pushed
up by a distance equal to 0.61 {\AA}. In the mean time, both adatoms get
displaced along [0$\bar1$0] by 0.5 {\AA} and 0.2 {\AA}, respectively.
Reconstruction in deeper silicon layers is not significant and it
ranges around $0.01 - 0.06$ {\AA} only. This result agrees well with the
fact that Cs adatoms do not alter the silicon surface very much as
was also suggested by Kim et al.\cite{Kim91} The reconstruction of
the surface leads to symmetric dimers with dimer lengths of 2.46 {\AA}
in excellent agreement with the experimental value\cite{Mangat} of
$2.48 \pm 0.06$ {\AA}. The symmetrization of silicon dimers by Cs
adsorption was found in agreement with various other experimental
results\cite{Hamamatsu,Meyerheim} in contrast to the high-resolution
core-level spectroscopy and angle-resolved valence-band spectroscopy
of Chao et al.\cite{Chao1,Chao3} who argued that the Si dimers stay
asymmetric. Abukawa et al.\cite{Abukawa98} studied the surface by
means of core-level photoemission for the covered surface
Si(001)2$\times$1-Cs and found that the surface dimers are symmetric
and excluded the results of Chao et al.\cite{Chao1,Chao3}

\subsection{Electronic Properties}

\subsubsection{Energy Bands}

\begin{figure}[h]
\epsfig{file=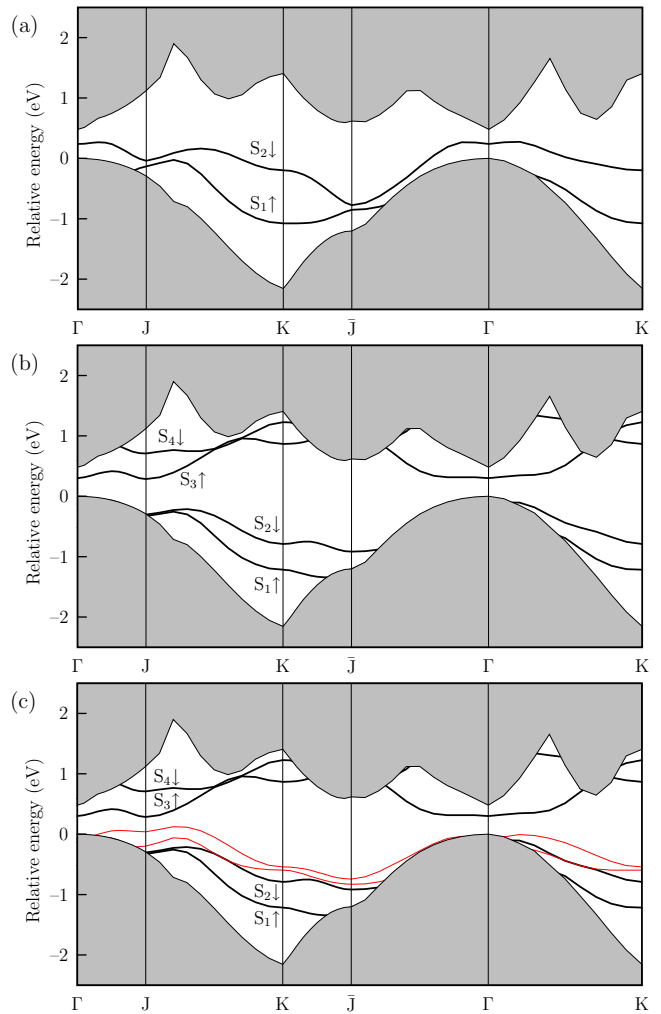,width=8.5cm,clip=true} \caption{Electronic band
structure of the Si(001) 2$\times$1 surface for (a) 0.5 ML Cs
coverage and (b) 1 ML Abukawa model. Projected bulk continuum is
shown by the shaded region. (c) Same substrate as in (b)
reconstructed due to adsorption of 1 ML Cs (thick lines), but the Cs
adatoms are taken away (thin lines). \label{figure3}}
\end{figure}

In Figure~\ref{figure3} we show our calculated band configurations
for the lowest energy structures of the 0.5 and 1 ML covered Si(001)
surface. For the clean surface we have two surface states in the
fundamental gap, namely S$_1$ and S$_2$, and the S$_2$ state usually
appears above the Fermi level. When the Cs atom gets adsorbed on
{\ih} site for the half ML case, its effect will be the shifting of
this state downward towards the valence bands, crossing the Fermi
level and causing a metallic behavior due to the partial filling of
the S$_2$ dangling bond band (Fig. 3(a)).

For the full ML coverage with adsorption of Abukawa type ({\hp}),
the dispersion shows a semiconducting surface as expected, and in
agreement with ARUPS results of Enta et al.\cite{Enta1} We have a
direct gap at $\Gamma$ with a band gap value of 0.30 eV (Fig. 3(b)).
This semiconducting nature of the surface might be due to the
saturation of the dimer dangling bonds. Benemanskaya et
al.\cite{Benemanskaya} expected that at 1 ML coverage, the adsorbed
surface will be semiconducting with two surface state bands, S$_3$
and S$_4$, appearing in the band gap below the bottom of the
conduction band which is in good agreement with our theoretical
results.

\begin{figure*}[ht]
\epsfig{file=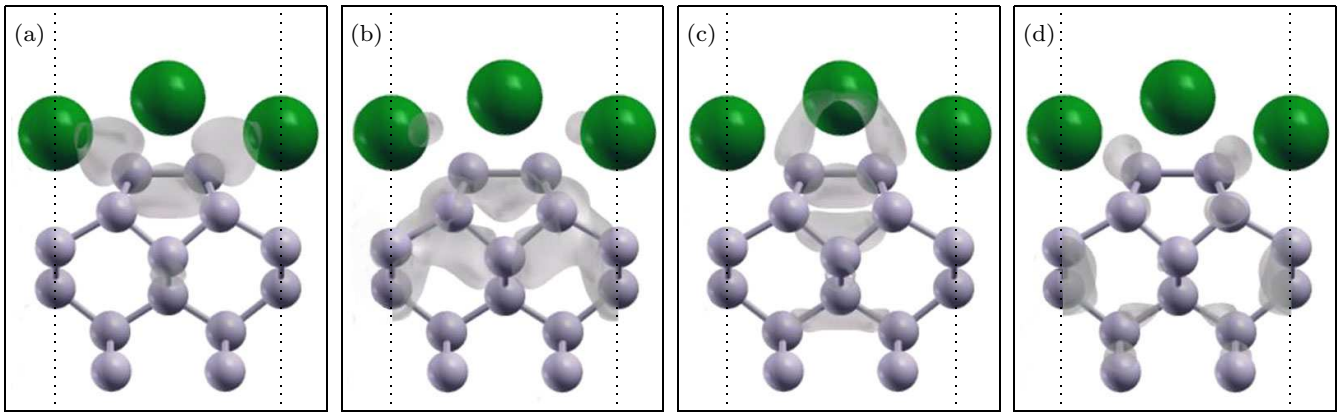,width=17.8cm} \caption{{\ih}-{\ip} model 3D
charge density distributions for $S_1$ and $S_2$ bands at high
symmetry points K and J in the surface Brillouin zone. Dashed lines
show the SUC boundaries. (a) $S_1$ at K, (b) $S_2$ at K, (c) $S_1$
at J and (d) $S_2$ at J. \label{figure4}}
\end{figure*}

The occupied band S$_1$ represents the $\pi$-like silicon dangling
bond surface state, while S$_2$ corresponds to the $\pi^*$-like Si
dangling bond state, which are shown in Figure~\ref{figure3}. In
order to study the nature of these states we calculated the 3D
charge density distributions for these occupied surface states at 1
ML coverage at the K and J high symmetry SBZ points. These plots are
given for the surface state S$_1$ in Figures~\ref{figure4}(a) and
\ref{figure4}(c), and for S$_2$ in \ref{figure4}(b) and
\ref{figure4}(d) at K and J points, respectively. The electronic
character of $\pi$ and $\pi^*$ Si dangling bond surface states show
partial contribution from the adsorbed Cs atoms which can readily be
seen in the Figures~\ref{figure4}(c) and \ref{figure4}(d) at the J
point of the SBZ. At J point, in Figure~\ref{figure4}(c), the
contribution to surface state S$_1$ comes from the upper Cs adatom
while it is vice versa for S$_2$, as seen in
Figure~\ref{figure4}(d). At K point the composition of the surface
states S$_1$ and S$_2$ show more contribution from the lower Cs
adatom.

The unoccupied low-lying surface states originate mainly because of
the adsorbed Cs atoms. They show a hybrid character as a combination
of Cs atomic $s$-orbital and dimer Si dangling bond orbitals.


The adsorption of Cs atoms causes a substantial shift in the
energies of surface states S$_1$ and S$_2$ towards bulk valence band
continuum over the whole SBZ as a result of the increase in the
binding energies. In the case of 1 ML coverage
Figure~\ref{figure3}(b) shows maximum change in the dispersions for
S$_1$ at J and for S$_2$ at K. Other significant deviations appear
at $\bar{\rm J}$ and J for S$_1$ and S$_2$, respectively. At K and
$\bar{\rm J}$ points S$_1$ and S$_2$ get closer when no Cs present.
This implies Cs-Cs electronic interaction is dominant at K and
$\bar{\rm J}$. At J point, on the other hand, Cs-Si interaction
seems to be more pronounced. This is because when Cs atoms are
removed the electron-electron interaction around dimer Si atoms
becomes more repulsive.

The shift in the energy values for the surface states S$_1$ and
S$_2$ comes mainly from the lowering of kinetic energy rather than
potential contribution to eigenenergies for these surface states.
Ishida et al.\cite{Ishi} showed for K/Si(001) system that this
lowering of kinetic energy increases with increasing K adsorption.
This behavior can be seen as a charge localization in the vicinity
of dimer Si and Cs atoms. Our results support the interpretation of
Soukiassian et al.\cite{soukiassian} of these shifts as being due to
Si 3$p$--Cs 6$s$ hybridization.

\subsubsection{Nature of the Cs-Si Bond}

\begin{figure}[b] \epsfig{file=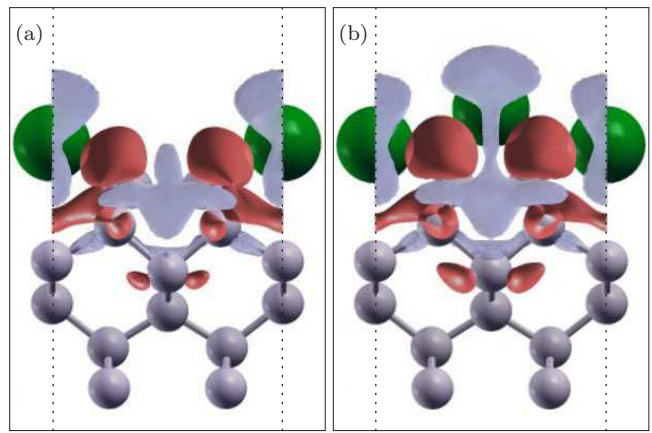,width=8.6cm,clip=true}
\caption{Electronic charge density difference 3D plots for 0.5 ML
and 1 ML coverages. Blue (bright) and red (dark) regions denote
charge depleted and charge accumulated zones, respectively.
\label{figure5}}
\end{figure}

Because the charge of Cs atom is much lower than that of Si bulk,
the total charge density of the adsorbed surface will show a bare Cs
atom, which makes the total charge density not meaningful in
studying the charge distribution for the adsorbed surface. Hence we
can not decide on the ionicity or covalency from the total charge
density picture.\cite{Ishi,kaba} This leads us to investigate the
difference in charge distributions instead. Our results for the
charge difference are very similar to those obtained
previously.\cite{Ishi,kaba} In Figure~\ref{figure5}(a) and (b), we
show 3D plots of the difference in charge distributions for the half
and full coverage adsorption, which is defined as
\begin{equation}
\delta\rho(r)=\rho_\mathrm{Cs/Si(001)}(r)-\rho_\mathrm{Si(001)}(r)-\rho_\mathrm{Cs}(r)
\end{equation}
where $\rho_\mathrm{Cs/Si(001)}(r)$ corresponds to the total charge
density of the covered surface, $\rho_\mathrm{Si(001)}(r)$ is the
total charge density of the Si substrate with the same atomic
arrangement as that of the covered surface, and
$\rho_\mathrm{Cs}(r)$ is the total charge density of the isolated Cs
atom.
We notice from Figure~\ref{figure5} that the strongest bonding
charge is in the plane which includes the silicon dimers and not in
the plane containing the Cs atoms. We also see that the charge
depletion for Cs atoms adsorbed on the surface takes place mostly
along [001] direction. Another major depletion takes place in
between the dimer atoms. This depletion results in favor of
accumulation of the charge between the dimer and Cs atoms.
Similarly, the depletion of charge around the Cs atom was found to
be mainly in the dimer plane rather than in the Cs adlayer. The
charge depletion around Cs atom in the vacuum direction is a little
bit shrunk for 1 ML coverage which comes as a sign of weaker Cs-Si
interaction and a smaller polarization of the bond. This result
suggest that we have polarized covalent bonding as was suggested
before by Ishida et al.\cite{Ishi} The bond becomes less polarized
as Cs coverage increases. Our theoretical Cs-Si bond lengths for the
0.5 and 1 ML cases being 3.58 {\AA} and 3.45 {\AA} are close to the sum of
the Cs (2.35 \AA) and Si (1.17 \AA) covalent radii, moreover they
compare well with the minimum bond length of 3.56 {\AA} for the CsSi
compound.\cite{Bus} The polarized covalent nature of the Cs-Si
bonding is also in agreement with the experimental results of
Soukiassian et al.\cite{soukiassian}

On the basis of these results we can now interpret the
symmetrization of dimers to be as result of saturation of dangling
bonds via hybridization. Since the Cs $6s$ orbital has a largely
spread wave function, Cs atom may interact with many neighboring Si
atoms resulting in symmetrization of the dimers at low coverages
such as 0.5 ML or even smaller, e.g., 0.3 ML as was shown
experimentally.\cite{Mangat} This may also explain the small
reconstruction that take place in the substrate which we believe is
due to the very limited charge transfer from the Cs atom to the
substrate.

\subsection{Work Function}

The work function is calculated as the difference between the vacuum
level and the Fermi energy, $E_{\rm F}$. The vacuum level is
determined from the self-consistent, plane-averaged potential,
$V_{\rm av}$, in the middle of the symmetrical slabs along the
direction perpendicular to the surfaces. It can be obtained from the
Poisson's equation,
\begin{equation}
{\partial^2\over\partial z^2}V_{\rm av}(z)=-4\pi\rho_{\rm av}(z)
\end{equation}
and is given by the relation,
\begin{equation}
V_{\rm av}(z) = -4\pi\int_z^\infty \!\!\! \rho_{\rm av}(z') \,z' \,
dz' + 4\pi z \int_z^\infty \!\!\! \rho_{\rm av}(z') \, dz'
\end{equation}
where
\begin{equation}
\rho_{\rm av}(z)={1\over A}\int\!\!\!\!\int_{A}
\vert\psi(x,y,z)\vert^2 \,dx\,dy\,.
\end{equation}

\begin{figure}[ht]
\epsfig{file=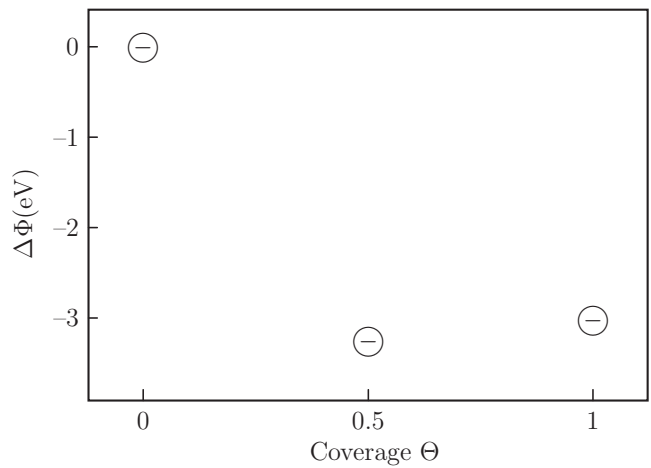,width=8.5cm,clip=true} \caption {Calculated
shifts in the work function as a function of the coverage with
respect to the clean surface\label{figure6}}
\end{figure}

The work function for clean Si(001)-2$\times$1 surface was reported
to be 4.9 eV by Abukawa et al.\cite{abukawa90} This agrees very well
with our theoretical value of 4.9 eV for the clean relaxed surface.
In Figure~\ref{figure6} we present our calculated values for the
work function shifts with respect to the clean surface, which are
3.26 eV and 3.02 eV for half and full coverages, respectively. The
lowering in the work function is more rapid between the clean
surface and the 0.5 ML coverage before it rises again by 0.24 eV at
the saturation coverage. Our calculated value for the lowering of
the work function for the full coverage compares well with the other
experimental values as shown in Table~\ref{table3}. This behavior in
the work function shift is consistent with experimental
results\cite{Enta1,Ort} for Cs/Si(001). It was found in most of the
experimental results that work function shifts reach to some maximum
value at a smaller coverage than that of saturation coverage before
it backs up a little to a value for the 1 ML. Since we found that
bonding between Cs atom and the silicon surface is polarized
covalent, we can not explain the lowering in the work function on
the basis of the classical Gurney picture\cite{Gurney} which is
based on the charge transfer and ideal ionic bonding systems. Hence,
we found that the model suggested previously by Ishida et
al.\cite{Ishi} might be more suitable to our results. In the initial
stages of Cs adsorption, the strong modification of charge
distribution around the Cs atoms will result in a polarized adlayer.
Adsorbed Cs atoms, thus, can be considered as dipole moments as a
result of charge depletion from the Cs adatom vacuum sides and the
charge accumulation between Cs-Si atoms. This signifies a
hybridization of Si $3p$ and Cs $6s$ which implies the saturation of
Si dangling bonds, leading to a rapid drop of the work function.
Even if the single dipole moments get weakened as a function of
increasing Cs coverage, the number of dipoles on the surface
increases making the overall dipole-dipole interaction significant.
This raises the potential energy barrier for outgoing electrons and
results in an increase in the work function by 0.24 eV as the
coverage goes from 0.5 ML to 1 ML.
\begin{table}%[h]
\caption{Work function change $\Delta\Phi$ (in eV) at the saturation
coverage with respect to the clean Si(001) $2\times1$ surface.
\label{table3}}
\begin{tabular}{c|c|c}
   \hline\hline && \\[-3mm]
   $\Delta\Phi$(eV) & Reference & Method \\
   \hline && \\[-3mm]
   \hspace{10pt} --3.02 \hspace{10pt} & \hspace{10pt} This work
   \hspace{10pt} & ab initio \\
   --3.11 & Ref. ~~[4] & ARUPS \\
   --3.4~~ & Ref. [35] & VBPS, CLPS \\
   --3.2~~ & Ref. [36] & EELS \\
   --2.8~~ & Ref. [37] & AES, LEED \\ \hline\hline
\end{tabular}
\end{table}

\subsection{Thermodynamic Stability of the Phases}

The thermodynamic stability for these coverages can be studied by
calculating the surface formation energy as a function of chemical
potential for each case. This allows one to compare the stability
of surface having different number of Cs adatoms.

\begin{figure}[htb]
\epsfig{file=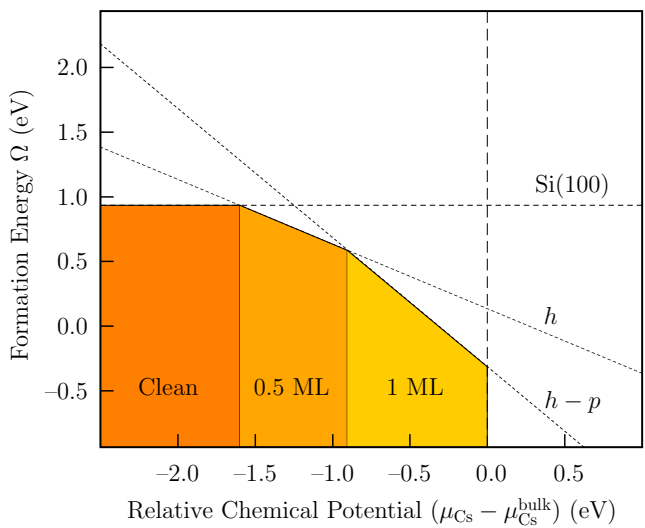,width=8.5cm,clip=true} \caption{The surface
formation energy as a function of the relative chemical potential of
Cs adsorbed on Si(001) surface per ($1\times1$) unit
cell.\label{figure7}}
\end{figure}

The surface formation energy\cite{qian} at 0 K for a slab containing
$n_{\rm Si}$ atoms of Si and $n_{\rm Cs}$ atoms of Cs adsorbed on
its surface can be written as
\begin{equation} \Omega=E_{\mathrm{Cs/Si(001)}}-n_{\rm Si} \,
\mu_{\mathrm{Si}} - n_{\rm Cs} \, \mu_{\mathrm{Cs}} \end{equation}
where $\mu_{\mathrm{Cs}}$ and $\mu_{\mathrm{Si}}$ are the chemical
potentials of Cs and Si, respectively. $E_{\mathrm{Cs/Si(001)}}$ is
the total energy for the covered surface as obtained by the
self-consistent ab initio calculations. In the usual adsorption
experiments using gaseous Cs, $\mu_{\mathrm{Cs}}$ of gaseous Cs is
less than the corresponding value of bulk, and it varies with
pressure and temperature. Thus surface chemical potential $\mu_{\rm
Cs}$ of Cs should be considered in some range such that it cannot
exceed the $\mu_{\rm Cs}^{\rm bulk}$. Taking the origin as the
chemical potential of bulk Cs, the calculated formation energies as
a function of $\mu_{\mathrm{Cs}}$ were presented in
Figure~\ref{figure7}.

It was found that for $(\mu_{\rm Cs}-\mu_{\rm Cs}^{\rm bulk}) <
-1.60$ eV the clean Si(001) surface has a lower formation energy
than that with any amount of Cs adsorbed on it. However, in the
range $-1.60$ eV~$< (\mu_{\rm Cs}-\mu_{\rm Cs}^{\rm bulk}) <
-0.90$~eV, the half coverage adsorption with a 2$\times$1 symmetry
is found to be the most stable structure. A transition to full
coverage with 2$\times$1 surface has occurred when $(\mu_{\rm
Cs}-\mu_{\rm Cs}^{\rm bulk}) > -0.90$ eV.

Since the half coverage is thermodynamically stable with a work
function difference close to that of the saturated surface, this
might be the reason why some experimental results were interpreted
as the half ML being the saturation coverage.

Our results show that after the completion of the half coverage with
all hollow sites occupied, if extra Cs atoms are added, they will
start filling the pedestal sites until a complete 1 ML is formed.
Thus, the saturation coverage for Cs/Si(001) system is 1 ML  and is
in agreement with the experimental
results.~\cite{Kim91,Lin,Mangat,Chao1,Chao2,Chao3,Benemanskaya,Hamamatsu}

\subsection{Conclusion}

We have performed an ab initio total energy calculation and geometry
optimization for a clean Si(001) surface and that with Cs overlayer
for 0.5 and 1 ML coverages on it. For the 0.5 ML coverage we have
found that Cs occupy the {\ih} site between dimer rows. The
adsorption sites for 1 ML coverage agree well with the Abukawa model
with {\hp} configuration. While our findings suggest that 0.5 ML is
thermodynamically stable, we found that 1 ML coverage is the
saturation coverage. The energy band spectra show metallic and
semiconducting surfaces for the half and full ML coverages,
respectively. The results for difference in charge upon Cs
adsorption suggests that the nature of the Cs-Si bonding is
polarized covalent. Our calculated values for the work function
difference agrees well with the existing experimental results.

\begin{acknowledgments}
This work was supported by T{\"U}B\.{I}TAK, The Scientific and Technical
Research Council of Turkey, Grant No. TBAG-2036 (101T058).
\end{acknowledgments}

% Create the reference section using BibTeX:
\bibliography{basename of .bib file}

\end{document}